\pdfoutput=1
\documentclass[%
 reprint,
 twocolumn,
 amsmath,amssymb,
 pra,
]{revtex4-2}

\usepackage{graphicx}                   % Include figure files
\usepackage{dcolumn}                    % Align table columns on decimal point
\usepackage{bm}                         % bold math
\usepackage{siunitx}                    % typeset units correctly
\usepackage{circuitikz}                 % draw circuits
\usepackage{lineno}                     % line numbers; use \linenumbers for the whole document
\usepackage{amsmath,amssymb}            % math
\usepackage[                            % set the page formatting
letterpaper,
]{geometry}

\makeatletter
\providecommand*{\diff}%
  {\@ifnextchar^{\DIfF}{\DIfF^{}}}
\def\DIfF^#1{%
  \mathop{\mathrm{\mathstrut d}}%
    \nolimits^{#1}\gobblespace}
\def\gobblespace{%
    \futurelet\diffarg\opspace}
\def\opspace{%
  \let\DiffSpace\!%
  \ifx\diffarg(%
    \let\DiffSpace\relax
  \else
    \ifx\diffarg[%
    \let\DiffSpace\relax
    \else
      \ifx\diffarg\{%
    \let\DiffSpace\relax
      \fi\fi\fi\DiffSpace}
\providecommand*{\Diff}%
  {\@ifnextchar^{\DDIfF}{\DDIfF^{}}}
\def\DDIfF^#1{%
  \mathop{\mathrm{\mathstrut D}}%
    \nolimits^{#1}\gobblespace}
\makeatother

\newcommand*\around{{\raise.17ex\hbox{$\scriptstyle\mathtt{\sim}$}}}

\usepackage[nomargin,inline,draft,author=]{fixme}       % use final instead of draft to remove all warnings
\fxusetheme{color}
\definecolor{fxwarning}{rgb}{0.8,0.0000,0.0000}
\FXRegisterAuthor{hm}{hdm}{hm}
\usepackage[colorlinks=false,bookmarks=false,citecolor=blue,urlcolor=blue]{hyperref} % pdflatex
\graphicspath{{figs}}                  % set the figures directory
\DeclareGraphicsExtensions{.pdf,.png}   % set the default image extensions

\begin{document}

% \preprint{APS/123-QED}

\title{Nonlinear dynamics in neuromorphic photonic networks: physical simulation in Verilog-A} % Force line breaks with \\
% \thanks{A footnote to the article title} %

\author{Hugh Morison}
 \email{hugh.morison@queensu.ca}
%  \altaffiliation[Also at ]{Physics Department, XYZ University.}%Lines break automatically or can be forced with \\
\author{Jagmeet Singh}%
\author{Nayem Al Kayed}
\author{A. Aadhi} % A Aadhi
\author{Maryam Moridsadat}
\author{Marcus Tamura}
\author{Alexander N. Tait$^1$}
\author{Bhavin J. Shastri}
\affiliation{%
Centre for Nanophotonics, Department of Physics, Engineering Physics, and Astronomy, Queen's University, Canada
}%
\affiliation{
$^1$Department of Electrical and Computer Engineering, Queen's University, Canada
}%


% \affiliation{
%  Third institution, the second for Charlie Author
% \affiliation{%
%  Authors' institution and/or address\\
%  This line break force with \textbackslash\textbackslash
% }%

% \collaboration{CLEO Collaboration}%\noaffiliation

\date{January 18, 2024}

\begin{abstract}
    Advances in silicon photonics technology have enabled the field of neuromorphic photonics, where analog neuron-like processing elements are implemented in silicon photonics technology.
    Accurate and scalable simulation tools for photonic integrated circuits are critical for designing neuromorphic photonic circuits.
    This is especially important when designing networks with recurrent connections, where the dynamics of the system may give rise to unstable and oscillatory solutions which need to be accurately modelled.
    These tools must simultaneously simulate the analog electronics and the multi-channel (wavelength-division-multiplexed) photonics contained in a photonic neuron to accurately predict on-chip behaviour.
    In this paper, we utilize a Verilog-A model of the photonic neural network to investigate the dynamics of recurrent integrated circuits.
    We begin by reviewing the theory of continuous-time recurrent neural networks as dynamical systems and the relation of these dynamics to important physical features of photonic neurons such as cascadability.
    We then present the neural dynamics of systems of one and two neurons in the simulated Verilog-A circuit, which are compared to the expected dynamics of the abstract CTRNN model.
    Due to the presence of parasitic circuit elements in the Verilog-A simulation, it is seen that there is a topological equivalence, but not an exact isomorphism, between the theoretical model and the simulated model.
    The implications of these discrepancies for the design of neuromorphic photonic circuits are discussed.
    Our findings pave the way for the practical implementation of large-scale silicon photonic recurrent neural networks.

% \begin{description}
% Secondary publications and information retrieval purposes.
% You may use the \texttt{description} environment to structure your abstract;
% use the optional argument of the \verb+\item+ command to give the category of each item.
% \end{description}

\end{abstract}

\maketitle

% \linenumbers
%\tableofcontents

\section{Introduction}
    \label{sec:intro}
    % processing that reflects the dynamical nature of both neural processing and the physical environment.

    As computers based on the von Neumann architecture\textemdash computers with separate processor and memory (CPUs, GPUs, and TPUs~\cite{wangBenchmarkingTPUGPU2019})\textemdash get faster, and as the datasets used to train machine learning models get larger, training time for state of the art models is increasingly consumed by memory access operations rather than computations \cite{backusCanProgrammingBe1978,williamsWhatNextEnd2017}.
    Neuromorphic engineering offers the potential to surpass this von Neumann bottleneck faced by digital processors, but also importantly may enable new application regimes by directly modeling neurons at the hardware level \cite{haslerFindingRoadmapAchieve2013,berggrenRoadmapEmergingHardware2021,daviesLoihiNeuromorphicManycore2018,merollaMillionSpikingneuronIntegrated2014,benjaminNeurogridMixedAnalogDigitalMultichip2014}.
    There has been much recent interest in the accurate modelling and control of complex systems of nonlinear differential equations.
    Notably, the Nobel Prize in Physics in 2021 was awarded to Parisi, Hasselmann, and Manabe for contributions to the understanding of nonlinear dynamical systems \cite{ComplexPredictable2022}.
    Moreover, the team at DeepMind has shown that neural networks and machine learning techniques can be used to tackle the real-time prediction and control of fast and highly nonlinear dynamical systems such as the magnetic confinement and shaping of plasma in nuclear fusion reactors \cite{degraveMagneticControlTokamak2022}.
    Neuromorphic circuits have a behavioural repertoire including neural control algorithms, as well as emulation of differential equations.
    Analog silicon photonic circuits have emerged as a promising platform to implement neuromorphic architectures with high bandwidth and low latency operation \cite{thomsonRoadmapSiliconPhotonics2016,prucnalNeuromorphicPhotonics2017,limaProgressNeuromorphicPhotonics2017,delimaMachineLearningNeuromorphic2019,shastri_photonics_2021}.
    While analog systems offer many advantages, they are sensitive to a wide variety of physical and environmental parameters so high accuracy simulation relies on physically detailed models that are validated with experimental observations.
    It was first shown in \cite{taitNeuromorphicPhotonicNetworks2017} how a network of silicon photonic neurons can be modelled as a Continuous-Time Recurrent Neural Network (CTRNN).

    The CTRNN is a dynamical model that can be applied to problems including differential equation emulation and neural control.
    CTRNNs are able to take on a wide range of dynamics by programming their interconnection weights and the nonlinear behaviour of each neuron.
    The most widespread example of a CTRNN is the Hopfield Network \cite{hopfieldNeuralNetworksPhysical1982,ramsauerHopfieldNetworksAll2020}.
    Hopfield networks naturally minimize a programmable energy function, and it has been shown that many control and optimization problems can be mapped to such an energy function \cite{hopfieldComputingNeuralCircuits1986,kennedyNeuralNetworksNonlinear1988a,wenReviewHopfieldNeural2009}.
    Hopfield networks are considered suitable candidates for a hardware implementation of optimization problems, but the length of the convergence time for such problems is directly related to the system's feedback latency.
    % talk about other digital neuromorphic circuits like Jagmeet does in Ch.1?

    Silicon photonics achieves massively parallel, low latency transmission through the use of passive optical waveguides and wavelength division multiplexing (WDM) \cite{goodmanFaninFanoutOptical1985,taitBroadcastWeightIntegrated2014,perezMultipurposeSiliconPhotonics2017}.
    Additionally, it supports high-bandwidth processing via an electro-optic modulation scheme using active modulators with demonstrated bandwidth \qty{>50}{\GHz} \cite{chrostowskiSiliconPhotonicsDesign2015,chanCband67GHz2022}.
    %motivation for optics is clear -- we need a simulation tool
    % cite the recommended simulation types from Thomas paper?
    Figure~\ref{fig:photonic_neuron} depicts a schematic of the silicon photonic neuron model investigated in this work.
    Silicon photonic neurons using this architecture have been fabricated and tested, implementing both feed-forward and recurrent neural networks \cite{zhang_system--chip_2024,Marquez_2023,huang_silicon_2021,taitNeuromorphicPhotonicNetworks2017}, although fully-integrated experimental demonstrations of these systems, and fast iteration of system design, remain challenging, in part due to the lack of a simulation platform that can replicate the dynamics of the on-chip system to predict the behaviour of networks of neurons including the effect of the electrical parasitics involved in the experimental set up.

    Previous work has developed a set of Verilog-A-based models of photonic components composing the electro-optic transfer function of the photonic neuron, and fit the parameters of these models using experimental measurements \cite{singhNeuromorphicPhotonicCircuit2022}.
    We believe that the incorporation of such models in established electronics simulators will enable large scale co-simulation of the electronics and photonics that are faithful to the underlying physics and account for parasitic elements of on-chip implementations.
    In this paper, we will use the previously developed photonic Verilog-A models \cite{singhNeuromorphicPhotonicCircuit2022} to simulate the physical behaviour of networks of neurons.
    This work will demonstrate important neural dynamics in the simulated system and demonstrate the deviations between these simulated dynamics and the expected behaviour of an abstract CTRNN.
    We begin by reviewing the dynamics of analog neural networks and the argument for the isomorphism between the CTRNN model and a photonic implementation of such a system.
    We then demonstrate these neural dynamics in a simulated version of a photonic neural network that incorporates all the parasitics present in experimental systems.

\section{Dynamics of continuous-time recurrent neural networks}
    \label{sec:theory}

    \subsection{Linear Feedforward Dynamics}
        \label{sec:linear-dynamics}
        
        A CTRNN is a neural network with neurons modelled by differential equations, such that a network of these neurons can be treated as a dynamical system.
        Neurons are typically arranged in a layer consisting of $N$ neurons, where all neurons have the same input vector $\vec{x}$ and share the same structure, but are independently parameterized.

        The most common model of a continuous neuron can be described as a leaky integrator\textemdash or a low-pass filter (LPF)\textemdash that integrates the weighted sum of its pre-synaptic connections.
        The evolution of the internal state of the $i$-th LPF neuron $s_i$ with time constant $\tau_i$, bias $b_i$, and weight vector $\bm w_i$, is described in Eq.~(\ref{eq:LPF_single_neuron}) where $\dot{s_i}$ indicates the time derivative of the state.

        \begin{align}
            \label{eq:LPF_single_neuron}
            \tau_i \dot{s_i}(t)&=-(s_i(t) - b_i) + \bm{w}_{i} \cdot \bm{x}(t) \\
            {y_i}(t)&=\sigma_i \left[s_i(t)\right] \nonumber
        \end{align}
        %which, in the absence of any input, is given by $s_i^*(t) = b_i$; the neuron state is always attracted to its bias point. In the case of external inputs, the fixed point of the neuron state is shifted by the magnitude of the dot product of the input vector and the weight vector
        The activation function $\sigma_i$ defines the nonlinear relationship between the neuron state and its postsynaptic output $y_i$.
        For biological plausibility, this function is typically taken to be a monotonic function such as a sigmoid.

        A layer of neurons can be treated as a system of equations by vectorizing Eq.~(\ref{eq:LPF_single_neuron}).
        The scalar parameters become vectors and a fully-connected matrix $\text{W}_{x}$ is formed from the set of weight vectors $\left\{ \bm w_i \right\}$.
        We use the subscript $x$ to indicate the incoming signal that the matrix operates on.
        The system of equations representing a layer of neurons is then

        \begin{align}
        \label{eq:LPF_layer}
            \tau \dot{\bm s}(t) &= - (\bm s(t) - \bm b) + \text{W}_{x} \cdot \bm{x}(t) \\
            \bm{y}(t) &=\sigma \left[\bm s(t)\right], \nonumber
        \end{align}
        where we have assumed that all neurons in the layer have identical activation functions $\sigma$ and time constants $\tau$.
        As linear first order differential equations with negative linear coefficients, this system can be seen to evolve toward its stable fixed point
        \begin{align}
        \label{eq:ff_fixed_point}
        &\bm{s}^* = \text{W}_{x} \cdot \bm{x} + \bm b \nonumber \\
        &\bm y^*=\sigma \left[ \text{W}_{x} \cdot \bm{x} + \bm b \right].
        \end{align}
        This is easily seen to be a stable system for all parameter values with exponential convergence of the state toward its fixed point, governed by the response time of the neurons $\tau$.
        Note that the fixed point remains valid even for a non-stationary input $x$ that varies at a timescale much less than $\tau$ (this will be seen in Sec.~\ref{sec:fanin}).
        While the output of the system is a nonlinear transformation of the state, if it is connected to a subsequent layer of neurons, this next layer still only has linear dynamics with respect to the output variable $\bm y$.
        This model can be seen as a continuous-time version of a dense feedforward neural network.
        For sufficiently large $N$, a feedforward network can act as a universal function approximator \cite{HORNIK1989359} but without recurrent connections it cannot represent systems of coupled equations with feedback effects.

    \subsection{Nonlinear Feedback Dynamics}
        \label{sec:nonlinear-dynamics}
        While the feedforward neuron is ensured to be stable, the dynamics of such a system become much more rich through the introduction of feedback connections from the set of post-synaptic outputs of a layer, $\left\{ y_i \right\}$, to all the pre-synaptic inputs of that same layer through a coupling matrix $\text{W}_{y}$.
        In this case, there is a subsequent term added to Eq.~(\ref{eq:LPF_layer}) that introduces a nonlinear dependence on the current state of the layer into each neuron contained in the layer.
        The layer is now written as a system of coupled equations,
        \begin{alignat}{2}
        \label{eq:LPF_single_neuron_fb}
            \tau \dot{\bm s}(t) &= - &&(\bm s(t) - \bm b) + \text{W}_{x} \cdot \bm{x}(t) + \text{W}_{y} \cdot \bm{y}(t) \nonumber \\
            &= - &&(\bm s(t) - \bm b) + \text{W}_{x} \cdot \bm{x}(t) \nonumber \\
            &&&+ \text{W}_{y} \cdot \sigma \left[\bm{s}(t)\right].
        \end{alignat}
        % Note the coupling term $\boldsymbol{W}_{y} \vec{\sigma} \left[\vec{s}(t)\right]$.

        The capability of the system is now greatly increased; it has been shown that a CTRNN of the form of Eq.~(\ref{eq:LPF_single_neuron_fb}) can act as a universal approximator of not just any function, but of any smooth dynamics \cite{beerDynamicsSmallContinuousTime1995,liApproximationDynamicalTimevariant2005}.
        It is therefore unsurprising that it is no longer possible to seek closed form solutions for the trajectory of the system in the general case, but it can be fruitful to investigate the fixed points and stability in specific well-defined cases.
        In particular, we will discuss the stability and bifurcations in the parameter space of single-neuron and two-neuron systems.

        In this paper, unless otherwise specified, we will assume that all neurons have a sigmoidal activation function that can be parameterized with a scale factor $\alpha$, a steepness $\beta$, and a baseline activation level $\gamma$:
        \begin{equation}
        \label{eq:activation_sigmoid}
            \sigma[s]=\frac{\alpha}{1+\exp{\left(-\beta (s-s_0)\right)}}+\gamma
        \end{equation}
        The variable $s_0$ is used to centre the activation function around a non-zero activation point.
        It is important to note that the forms of Eq.~(\ref{eq:LPF_single_neuron_fb}) and Eq.~(\ref{eq:activation_sigmoid}) are different from other work on this photonic neuron architecture which has used Mach-Zehnder Modulators and AC coupling between the BPD and modulator \cite{taitNeuromorphicPhotonicNetworks2017}.
        The equations we have chosen more closely match the system under study but give rise to inelegant solutions, so we rely on numerical methods to solve the equations presented in the following sections.
        In many cases, we care about the dynamics of the output variable $y$ rather than the state variable $s$ (because it corresponds to a more readily observable variable\textemdash the optical power in a channel) but this can always be computed using the chain rule,
        \begin{equation}
        \label{eq:chain_rule}
            \dot y_i \equiv\frac{\diff y_i}{\diff t} = \frac{\diff y_i}{\diff s_i}\frac{\diff s_i}{\diff t} = \sigma_i' \left[s_i\right] \dot s_i.
        \end{equation}

        \subsubsection{Single Neuron Networks}
            \label{sec:one-neuron}
            If we seek the fixed points to Eq.~(\ref{eq:LPF_single_neuron_fb}) in the case of a single neuron with a single scalar input $x$, using the activation function in Eq.~(\ref{eq:activation_sigmoid}), we have
            \begin{alignat}{2}
                \label{eqn:cusp_steady}
                0&=&&- (s^*- b) + x + {W}_{F} \sigma \left[s^*\right] \nonumber \\
                % 0 &= - W_F\gamma s^3 + (W_F\beta-1) s + W_F\alpha + b + \boldsymbol{W}_{x} \cdot \vec{x}
                &=&&-s^* + b + x \nonumber \\
                &&&+ W_F \left( \frac{\alpha}{1+\exp{\left(-\beta (s^*-s_0)\right)}} + \gamma \right),
            \end{alignat}
            where we have used a feedback weight $\text W_y=[W_F]$ and input weight $\text W_x=[+1]$.
            This equation defines the nullcline for a single neuron.
            We can't readily solve for $s^*$ analytically but we can note certain properties in (\ref{eqn:cusp_steady}).
            Namely, this equation is the sum of a negative linear function of $s^*$, some constants, and the sigmoid.
            Since the sigmoid is an odd function, it can be approximated by a cubic, and solving for the fixed point $s^*$ would be a matter of solving the resulting cubic equation.
            Under this cubic approximation, there are three roots; one of them is guaranteed to be real, while the other two may be imaginary.
            We can interpret the imaginary roots to be non-physical solutions and the real roots to be the true fixed points.
            Whether the solutions are real or imaginary depends on the values of the parameters $\alpha$, $\beta$, $\gamma$, and the weight $W_F$.
            In terms of the neural dynamics, this means that by varying the feedback weight, we can achieve between one and three different fixed points.
            In the case of three fixed points, only two will be stable while the third will be an unstable point.
            It has been shown in \cite{taitNeuromorphicPhotonicNetworks2017} how the observation of a transition from one stable fixed point (monostability) to two stable fixed points (bistability) is equivalent to a demonstration of the cascadability of the photonic neuron.

        \subsubsection{Two Neuron Networks}
            \label{sec:two-neurons}
            Networks containing two neurons are the simplest testbed for demonstrating the coupled dynamics of the system.
            See \cite{beerDynamicsSmallContinuousTime1995} for a comprehensive overview of CTRNN dynamics; depending on the programmed weight matrix, two neuron networks can have up to 9 qualitatively different phase portraits.
            Here, we will investigate two canonical examples of neural circuits that give rise to a stable system and an oscillatory system, respectively:
            %\begin{multicols}{2}
            \begin{equation}
            \label{eq:wta_matrix}
                \text W_{WTA} = \begin{bmatrix}
                                +1 & W_{inh} \\
                                W_{inh} & +1
                            \end{bmatrix}
            \end{equation} \hspace{1pt} % \break
            \begin{equation}
            \label{eq:hopf_matrix}
                \text  W_{Hopf} = \begin{bmatrix}
                                W_F & +1 \\
                                -1 & W_F 
                            \end{bmatrix}
            \end{equation}
            %\end{multicols}

            Both matrices take the place of the recurrent matrix $\text {W}_{y}$ in Eq.~(\ref{eq:LPF_single_neuron_fb}).
            In the symmetric matrix Eq.~(\ref{eq:wta_matrix}), the positive self-feedback and the inhibitory mutual feedback ($W_{inh}<0$) give rise to the winner-take-all (WTA) system.
            The mutual inhibition guarantees that the neuron with the highest level of activation will suppress the activity of its counterpart while the positive self-feedback ensures that it maintains its own level of activity.
            From a dynamics perspective, there are two basins of attraction in the state space around the points where one neuron is at maximum activity and the other is at minimum activity.
            % \begin{align}
            %     \label{eqn:wta_fixed_pts}
            %     0 &= - (s_1 - b_1) + \sigma \left(s_1\right) + W_{inh} \sigma \left(s_2\right) \nonumber \\
            %     &= -s_1 + \frac{\alpha}{1+\exp{-\beta s_1}} + \frac{W_{inh} \alpha}{1+\exp{-\beta s_2}} + b_1 \\
            %     0 &= - (s_2 - b_2) + \sigma \left(s_2\right) + W_{inh} \sigma \left(s_1\right) \nonumber \\
            %     &= -s_2 + \frac{\alpha}{1+\exp{-\beta s_2}} + \frac{W_{inh} \alpha}{1+\exp{-\beta s_1}} + b_2
            % \end{align}

            The asymmetric weight matrix Eq.~(\ref{eq:hopf_matrix}) results in an unstable system.
            When the self-feedback parameterized by $W_F$ is too low, the neurons won't have the activity required to reinforce themselves and they will settle at their stable fixed points, but, as $W_F$ increases, there is a bifurcation point where this stable state gives rise to an oscillatory solution (a stable limit cycle).
            This is known as the Hopf bifurcation.
            % \begin{align}
            %     \label{eqn:hopf_limit}
            %     0 &= - (s_1 - b_1) + {W}_{F} \sigma \left(s_1\right) + \sigma \left(s_2\right) \nonumber \\
            %     &= -s_1 + \frac{W_F \alpha}{1+\exp{-\beta s_1}} + \frac{\alpha}{1+\exp{-\beta s_2}} + (W_F+1) \gamma + b_1 \\
            %     0 &= - (s_2 - b_2) + {W}_{F} \sigma \left(s_2\right) - \sigma \left(s_1\right) \nonumber \\
            %     &= -s_2 + \frac{W_F \alpha}{1+\exp{-\beta s_2}} - \frac{\alpha}{1+\exp{-\beta s_1}} + (W_F-1) \gamma + b_2
            % \end{align}

    \subsection{Silicon Photonic Implementation}
    \label{sec:sip-neuron}

    \begin{figure*}[htbp] % [b] [htbp], [H]
        \centering
        \includegraphics[width=\linewidth]{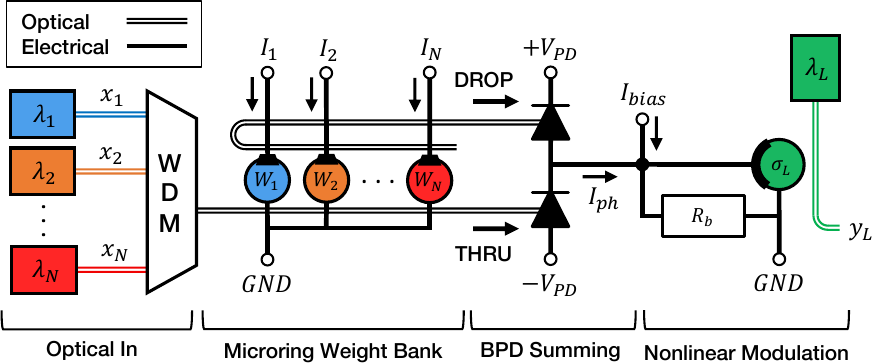}
        \caption{
            Schematic of a silicon photonic neuron with $N$ input channels.
            Each input signal $x_i$ is amplitude modulated onto an optical carrier with wavelength $\lambda_i$ (modulation not depicted) and then wavelength-division multiplexed (WDM) onto a single waveguide.
            Synapses are implemented with a bank of tunable add-drop microring filters.
            The labels $I_i$ represent current sources applied to in-ring heaters used to tune each filter and give its corresponding channel weight $W_i$.
            The photocurrent $I_{ph}$ generated by the balanced photodetector configuration is the result of the weighted sum across all the optical channels.
            Each neuron has electrical connections $\pm V_{PD}$ to bias the photodetectors.
            A PN junction-integrated microring modulator is used to apply a nonlinear activation function $\sigma_L$ to the weighted sum, modulating a signal $y_L$ onto a new optical carrier (here of wavelength $\lambda_L$, where $L\in{\lbrack 1, N \rbrack}$).
            The modulator is biased using a current source with current $I_{bias}$ to achieve a bias voltage $V_b=I_{bias} R_b$.
            % The connection points show where the integrated circuit connects to off-chip biasing and control electronics via bond pads that introduce parasitics; the bond pad model is described in the Appendix.
            The connection points show where bond pads introduce parasitics into the integrated circuit.
        }
        \label{fig:photonic_neuron}
    \end{figure*}

    The model of the photonic modulator neuron seen in Fig.~\ref{fig:photonic_neuron} is based on a microring modulator using electro-optic effects in PN junction-integrated silicon photonic waveguides.
    The nonlinearity $\sigma_i$ is achieved through the electro-optic transfer function of the microring modulator \cite{taitSiliconPhotonicModulator2019}.
    Each neuron is tuned to modulate a different channel in the WDM waveguide bus, and the spectral position of the nonlinear Lorentzian transmission spectrum of the modulator is tuned by the applied voltage generated by the photocurrent.

    In equations~(\ref{eq:LPF_single_neuron}\textendash \ref{eq:LPF_single_neuron_fb}), the state $s_i$ is therefore related to the voltage across this PN modulator $V_i$ and the bias $b_i$ is similarly related to a bias voltage $V_{b,i}$.
    The time constant of the system $\tau_i$ is governed by the bandwidth of the optical-electrical-optical conversion of the photonic neuron, which includes the PN modulator and the photoreceiver impedance.
    When operating in reverse bias (carrier-depletion mode), PN junction modulators act as a capacitive load with a cutoff frequency in the range of \qtyrange{40}{70}{\GHz} under typical bias conditions \cite{chrostowskiSiliconPhotonicsDesign2015}.
    When operated in forward bias (carrier-injection mode), we can use the same dynamical equation but the time constant is instead limited to \qty{<10}{\GHz} by the slower rate of carrier recombination in the PN junction \cite{chrostowskiSiliconPhotonicsDesign2015,liuReviewPerspectiveUltrafast2015}.
    In practice, a bandwidth of \qty{10}{\GHz} has been achievable with current technology \cite{delimaNoiseAnalysisPhotonic2020}.

    All the weighted inputs to the low pass filter come from the microring weight bank \cite{zhangSiliconMicroringSynapses2022} configuration seen in Fig.~\ref{fig:photonic_neuron}.
    The photocurrent generated by the balanced photodetector (BPD) pair is proportional to the difference in the optical power applied to each photodetector, where the optical power is the sum over the power in each individual channel, $P_j$.
    The power is split between the two photodetectors using the transmissions at the THRU-port ($T_{j}$) and DROP-port ($D_{j}$) of a tunable filter bank implemented by the cascaded microring resonators (MRRs).
    In this way, the combination of the MRRs and BPD generate a photocurrent proportional to the weighted sum $\bm{w}_i \cdot \bm{x}$.

    The resultant dimensional version of Eq.~(\ref{eq:LPF_single_neuron}) is given by
    \begin{align}
    \label{eq:LPF_single_neuron_dimensioned}
        \tau_i \dot{V_i}(t) = -& (V_i(t) - V_{b,i}) \nonumber \\
        &+ \sum_j \eta(D_{j}-T_{j})P_j(t),
    \end{align}
    where $\eta$ is the optical conversion efficiency (in units of V/W) which is related to the photodetector responsivity and the receiver impedance.

    A layer of neurons is formed by tuning each neuron's modulator to an individual channel wavelength.
    An isomorphism between a photonic neural network using this architecture and the CTRNN system given in Eq.~(\ref{eq:LPF_single_neuron_fb}) has been shown in \cite{taitNeuromorphicPhotonicNetworks2017} although the accuracy of this result benefitted from low bandwidth operation and the filtering and modulation (and therefore the dynamics) being located off-chip.
    Fully integrated networks in silicon photonics technology will follow the same dominant dynamics, but experimental demonstrations deviate from the perfect CTRNN model due to a number of on-chip effects.
    Parasitic capacitance (from bond pads) and inductance (from wire bonds), as well as reactive components of the biasing circuitry, introduce higher order terms with different time constants into Eq.~(\ref{eq:LPF_single_neuron_dimensioned}).
    Feedback delays (assumed to be negligible in all the presented equations) can also produce spurious dynamics, and the electro-optic activation functions implemented by microring modulators are not arbitrarily programmable, so they deviate from idealized sigmoid and ReLU functions used in machine learning.
    For these reasons, using dynamical analysis of Eq.~(\ref{eq:LPF_single_neuron_fb}) is good as an approximation and to understand the capabilities of dynamical network models, but is insufficient for predicting on-chip behaviour quantitatively.

\section{Verilog-A Neural Networks}
    \label{sec:results}
    Our previous work has shown a methodology for modelling a photonic neuron in Verilog-A \cite{singhNeuromorphicPhotonicCircuit2022}.
    Verilog-A is a hardware description language for analog devices that is compatible with a broad range of circuit simulators (e.g. SPICE, Spectre \cite{pratapReviewVariousAvailable2014,mcandrewSPICEModelingVerilogA2017}).
    Using Verilog-A enables accurate modeling of dynamic electro-optic effects and optimized co-simulation of the electronics present in photonic neuron circuits.
    %The optical signals in these models are represented as data structures containing real and imaginary electric field amplitudes along with their wavelength.
    We fit the parameter values of the electro-optic effect in PN microring modulators and the thermo-optic effect in N-doped microring resonators to experimentally determined values so that the models reflect the behaviour seen on chip.
    % There are active efforts to develop these models from theory to robustly account for all the relevant physics, but the experimental methodology allowed us to have simulated devices that better match the results seen on chips in lab.
    The results of this simulation methodology proved to accurately model the on-chip behaviour of a single photonic neuron.

    In this section, we examine the signal processing capability and dynamics of networks of simulated neurons in the presence of realistic parasitic components.
    All figures show a simplified circuit schematic, the full circuit includes bond pad capacitance, wire bond inductance, and a basic bias circuit; a detailed diagram is available in Appendix~\ref{sec:appendix-circuit-diagram}.
    Section~\ref{sec:fanin} will show feedforward time domain behaviour of this neuron model with multiple input signals to demonstrate excitatory and inhibitory fan-in.
    Section~\ref{sec:cascadability} will show the behaviour of this single neuron in a self-feedback circuit to demonstrate the indefinite cascadability of these neurons.
    Section~\ref{sec:verilog-dynamics} will examine the dynamical properties of the neuron through the analysis of a Hopf bifurcation in a two-neuron system.
    Section~\ref{sec:wta_dynamics} shows that by reconfiguring the weights, the same two-neuron circuit can give rise to WTA dynamics.

    \subsection{Signal Fan-in}
        \label{sec:fanin}

        \begin{figure*}[htbp] %[b]
            \centering
            \includegraphics[width=0.95\linewidth]{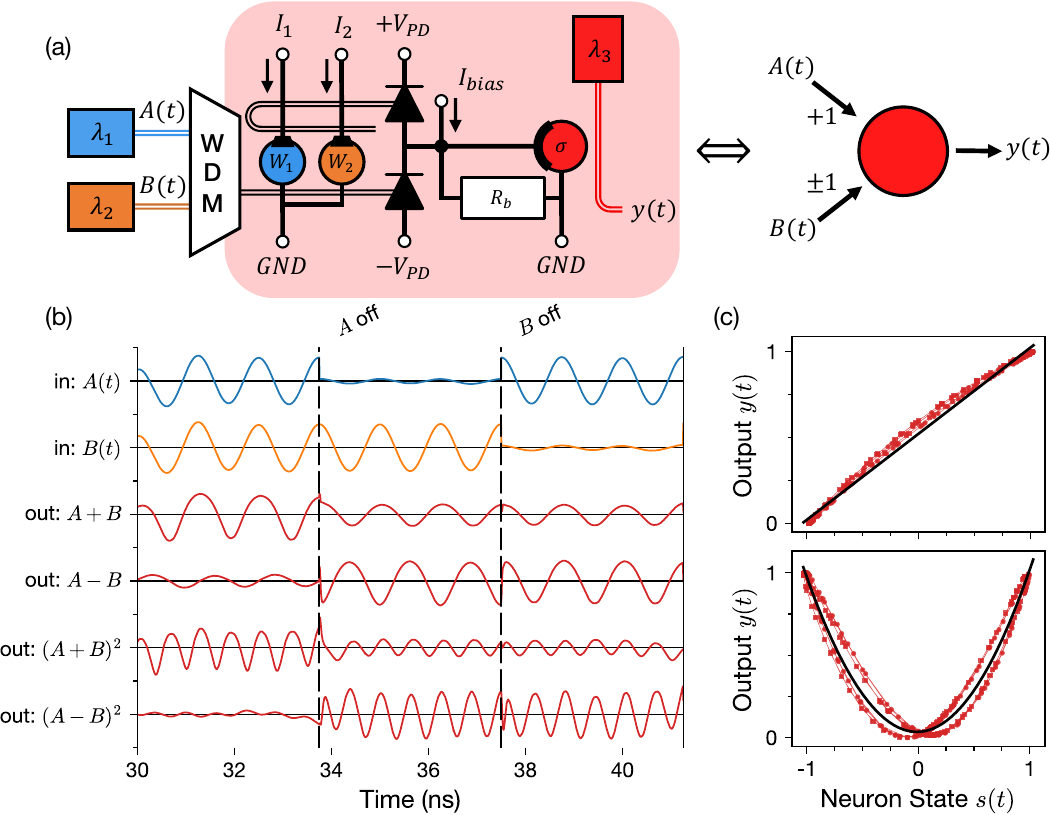}
            \caption{
                Two-channel addition and subtraction, and nonlinear activation functions in a simulated photonic modulator neuron.
                (a) Schematic of Verilog-A simulation and equivalent conceptual schematic.
                (b) Input and output optical signals for various circuit configurations.
                $A$ and $B$ are input signals at $\lambda_1$ and $\lambda_2$.
                Output signals on $\lambda_3$ demonstrate four different operations applied to the inputs, achieved by programming the weights and activation function.
                All signals have a high-pass filter with a cutoff frequency of \qty{400}{\MHz} applied to remove the DC component and are scaled to emphasize the qualitative aspects of the results.
                (c) Linear and quadratic activation functions obtained by plotting the output power of the neuron against the voltage of the junction.
                Circular markers are used for $A+B$ and $(A+B)^2$ while square markers used for $A-B$ and $(A-B)^2$.
                The black lines show fits to the data using a linear and quadratic functions.
            }
            \label{fig:burst_addition}
        \end{figure*}

        The feedforward fixed point of a single neuron described by Eq.~(\ref{eq:ff_fixed_point}) implies that the system should be able to perform addition and subtraction of two input channels by configuring its weights.
        We can then apply a linear or nonlinear activation function to this computed result by configuring the bias to tune the transfer function of the modulator.
        Figure~\ref{fig:burst_addition}(a) shows the schematic for this simulation.
        We follow the procedure developed in \cite{taitNeuromorphicPhotonicNetworks2017} and prepare two input signals $x_1=A(t)$ and $x_2=B(t)$ on optical carriers with wavelengths $\lambda_1$ and $\lambda_2$, both amplitude modulated at \qty{900}{\MHz}.
        In this work, the weighting is implemented by the tunable thermo-optic effect of the microring weight bank.
        We configure $W_1$ with a weight of $+1$ and $W_2$ with a weight of $\pm1$ to implement addition and subtraction of the two inputs; the detailed weight mapping procedure is described in Appendix~\ref{sec:appendix-weight-mapping}.
        The modulator then implements either a linear or a quadratic activation function to converts the signal onto a carrier of wavelength $\lambda_3$.
        %; the process of implementing various nonlinear activation functions is described in \ref{sec:appendix-activation-funcs}.

        Figure~\ref{fig:burst_addition}(b) shows the optical input signals $A(t)$ and $B(t)$ on the first two traces and the output signal $y(t)$ for a selection of different functions programmed into the neuron.
        The third trace $A+B$ shows the optical output of the neuron biased in the linear regime with $W_1=W_2=+1$.
        The fourth trace $A-B$ shows this same output after changing the weights to $W_1=+1$ and $W_2=-1$.
        The fifth $(A+B)^2$ and sixth $(A-B)^2$ traces show the same two cases for the weights, but with the neuron biased in the quadratic regime.
        Note the frequency doubling in the two quadratic cases.
            
        Figure~\ref{fig:burst_addition}(c) shows the activation functions obtained by plotting the output optical power against the junction voltage and the linear and quadratic fits to the data.
        Qualitatively, we can see good agreement between the simulated results and the linear and quadratic functions.
        There is some slight nonlinearity in the linear activation function, as well as some hysteresis evident in the quadractic activation function, but we still achieve a root-mean-square error of 3.8\% and 6.4\% for the linear and quadratic fits, respectively.
        
        These results demonstrate the substantial repertoire of behaviours of the feedforward photonic neuron.
        The MRR weights and BPD perform programmable excitatory and inhibitory fan-in of signals across multiple optical channels.
        The PN junction MRR modulator demonstrates a configurable nonlinear conversion back into the optical domain operating at high bandwidth.
        Taken together, these results show that the feedforward photonic neuron can be used as building block of a CTRNN.

        \begin{figure*}[htbp]
            \centering
            \includegraphics[width=0.95\linewidth]{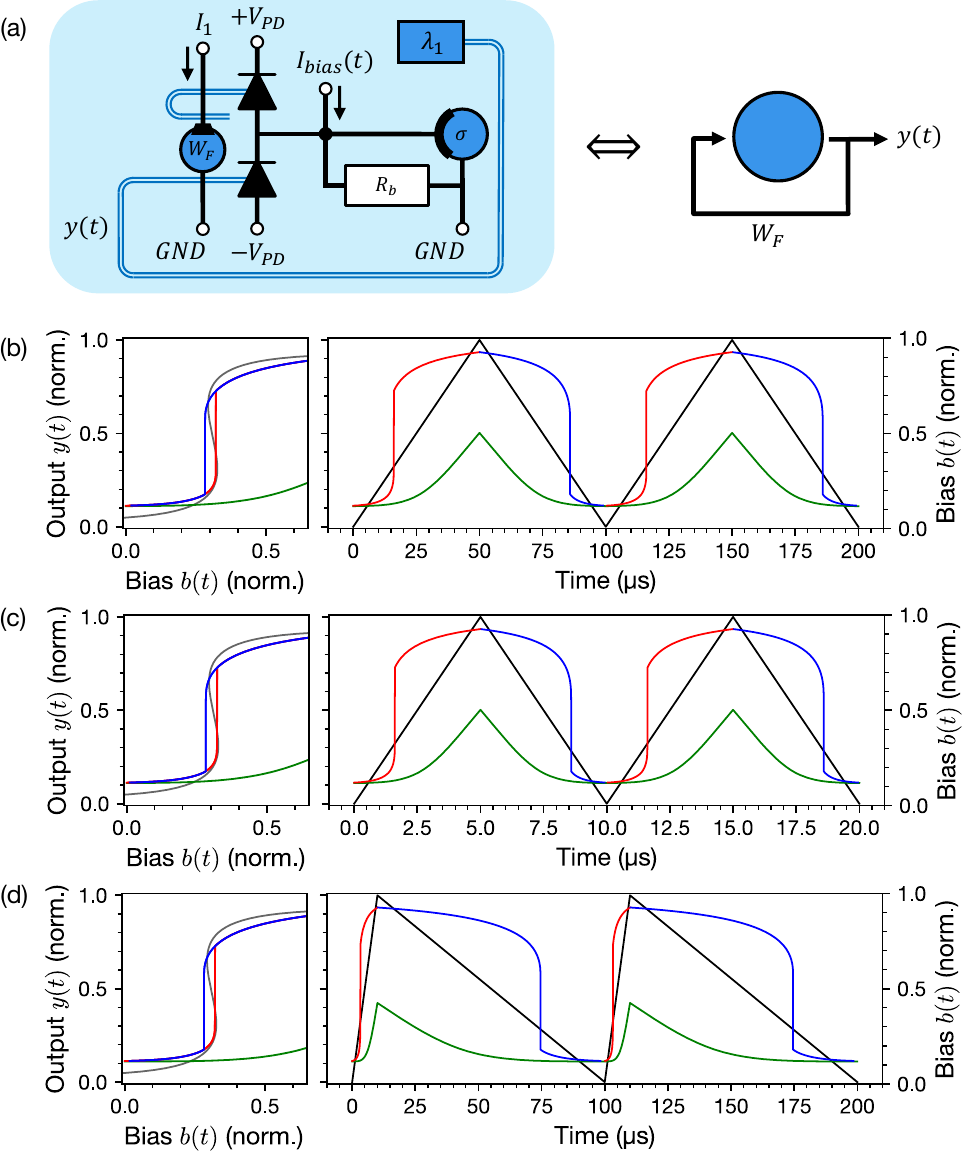}
            \caption{
                Bistability in single self-afferent neuron.
                (a) Schematic of simulated recurrent circuit and equivalent conceptual schematic.
                (b,c,d) Relationship between neuron bias $b(t)$ and output $y(t)$ for three different bias signals.
                The output $y$ on $\lambda_1$ when $W_F=0$ is shown in green, while the red and blue traces show the rising and falling edges of $y$ when $W_F=+1$.
                The $y$ vs. $b$ plots on the left shows the signals along with the system's nullcline (grey trace).
                The transient plots on the right shows the same signals along with the bias waveform (black trace).
                Optical powers are normalized using $y(t)=P(t)/P_{pump}$, the bias current is normalized using $b(t)=(I(t)-I_{min})/(I_{max}-I_{min})$.
            }
            \label{fig:casacadability_results}
        \end{figure*}

    \subsection{Cascadability}
        \label{sec:cascadability}
        Using the same neuron circuit as the feedforward case, connecting the output back to the input enables the self-feedback.
        For the following simulations, we configure the activation function to be a sigmoidal function, as described in Sec.~\ref{sec:nonlinear-dynamics}.
        We configure the feedback delay to be \qty{10}{\ps}, consistent with a \qty{1}{\mm} long feedback waveguide.
        Figure~\ref{fig:casacadability_results}(a) shows the simplified circuit diagram of this single neuron CTRNN and its conceptual representation.
        We use a pump of wavelength $\lambda_1$ and tune the neuron modulator to this wavelength.
        To induce bistability in this circuit, we use a time varying bias current and observe the optical output (post-modulation) on $\lambda_1$.
        Note that this is an atypical biasing configuration; static bias currents are typically used but in this simulation it is simpler to use a time-varying bias to highlight the dynamics that arise in the output optical power.

        For this simulation, it is important that the signal be of sufficiently low frequency to ensure that the observed dynamics are are a result of the self-feedback in the neuron and not a frequency-induced phase-shift between the input and output.
        We consider three different cases for the signal on $I_{bias}(t)$ to make this point.
        First, we use a \qty{10}{\kHz} triangular wave; next, we use an \qty{100}{\kHz} triangular wave to demonstrate that the effect is not frequency dependant; finally, we use a \qty{10}{\kHz} asymmetric triangular wave to demonstrate that the effect has no dependence on the shape of the driving function.
        In all cases, we vary the self-feedback weight $W_F$ from 0 to 1 to observe the effect of the feedback connection.

        %Fig.~\ref{fig:casacadability_results}(c) shows the results of this sweep in a 3-dimensional plot; each slice of the plot with a fixed $W_F$ is a transient simulation with the output plotted against the input.

        Figure~\ref{fig:casacadability_results}(b) shows the results of the simulation for the \qty{10}{\kHz} input signal.
        In the absence of feedback ($W_F=0$), the system is clearly monostable.
        When the feedback is introduced ($W_F=1$), we can see bistability indicated by the separation between the rising and falling edges of the output which follows the behaviour predicted by (\ref{eqn:cusp_steady}).
        Figure~\ref{fig:casacadability_results}(c) shows the results of the simulation for the \qty{100}{\kHz} input signal.
        Both the monostable and bistable behaviour are identical to the \qty{10}{\kHz} case, as expected.
        Again, the bistability is evident from the hysteresis of the bias-output relationship when $W_F=1$.
        Figure~\ref{fig:casacadability_results}(d) demonstrates how the time domain behaviour can vary considerably but the phase space representation again looks identical.

        As expected, in all cases, the solution is monostable for $W_F\approx0$.
        This is analogous to the feedforward results presented in Sec.~\ref{sec:fanin}; the output is just a nonlinear function of the neuron state defined by the neuron's activation function, and there is negligible frequency dependence.
        As $W_F$ is increased, the results imply a bifurcation point where the monostable response gives way to this bistable response.
        %The vector field is a plot of Eq.~(\ref{eq:LPF_single_neuron_fb}) using activation function Eq.~(\ref{eq:activation_sigmoid}).
        %The dashed line is the nullcline given by (\ref{eqn:cusp_steady}).
        %\hl{The results line up closely with the model and demonstrate the cascadability of the simulated neurons.}

        \begin{figure*}[htbp]
            \centering
            \includegraphics[width=\linewidth]{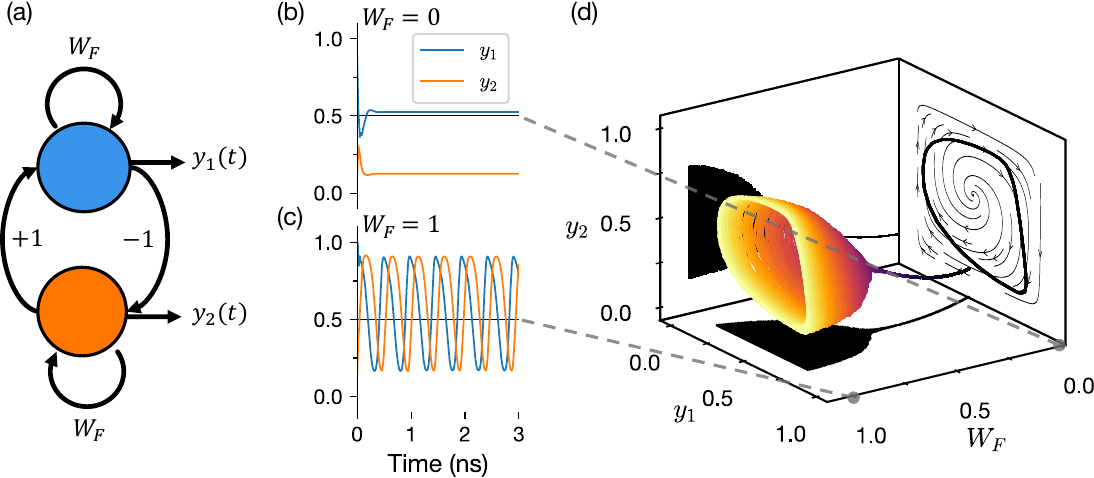}
            \caption{
                (a) Conceptual representation of the two-neuron circuit showing the weight values.
                (b,~c) Transient response of neuron outputs for $W_F=0$ and $W_F=1$, respectively.
                The outputs $y_i$ are given by the normalized output power from each modulator.
                (d) 3D bifurcation diagram of the Hopf system.
                The stability transitions from a stable fixed point to a stable limit cycle as $W_F$ is varied.
                Black shadows show the projections of simulated data onto the $y_i-W_F$ axes.
                The projection on the $y_1-y_2$ axes shows both the expected dynamics as a vector field, and the slice of the simulated data where $W_F=1$.
            }
            \label{fig:hopf-veriloga}
        \end{figure*}

    \subsection{Hopf Bifurcation}
        \label{sec:verilog-dynamics}

        The two dimensional system provides the simplest test circuit to demonstrate the interaction between neurons while still having easily interpreted phase portraits, as described in Sec.~\ref{sec:two-neurons}.
        In Fig.~\ref{fig:hopf-veriloga}(a), we depict the simple conceptual diagram of the network, with weights configured as shown in Eq.~(\ref{eq:hopf_matrix}).
        To observe the bifurcation in this circuit, we do not introduce any modulated input signal, as was done in the previous simulation.
        Instead, two lasers with wavelengths $\lambda_1$ and $\lambda_2$ are used as pumps and a time dependence of the modulation occurs due to the feedback dynamics of the circuit, governed by the weights and the activation functions.
        We observe the post-modulation signals over time while sweeping feedback weights $W_F$ from 0 to 1.

        The transient results for $W_F=0$ and $W_F=1$ are shown in Figures~\ref{fig:hopf-veriloga}(b) and~\ref{fig:hopf-veriloga}(c), respectively.
        In Fig.~\ref{fig:hopf-veriloga}(d), we see the bifurcation in the phase portrait of the system.
        When the feedback connection to each neuron is low, the system quickly decays to its single stable fixed point, no matter its starting position.

        Notice how the coordinate of the stable fixed point has a nonlinear relationship with $W_F$.
        Eventually, for large enough $W_F$, there is a bifurcation point where the fixed point evolves into a stable limit cycle.
        The amplitude of this limit cycle increases by increasing the strength of each neuron's self-feedback $W_F$.
        The vector field projected on the $y_1-y_2$ axes in Fig.~\ref{fig:hopf-veriloga}(d) is a plot of Eq.~(\ref{eq:LPF_single_neuron_fb}) using weight matrix Eq.~(\ref{eq:hopf_matrix}) and an activation function fit to the results.
        The simulated trajectory follows the dynamics of the model but there are notable deviations and asymmetries which arise from the combination of all the physical effects present in the simulation.
        This deviation in the simple two-neuron system depicts why physical simulation is necessary for accurate prediction of on-chip experimental results.

        \begin{figure*}[htbp]
            \centering
            \includegraphics[width=\linewidth]{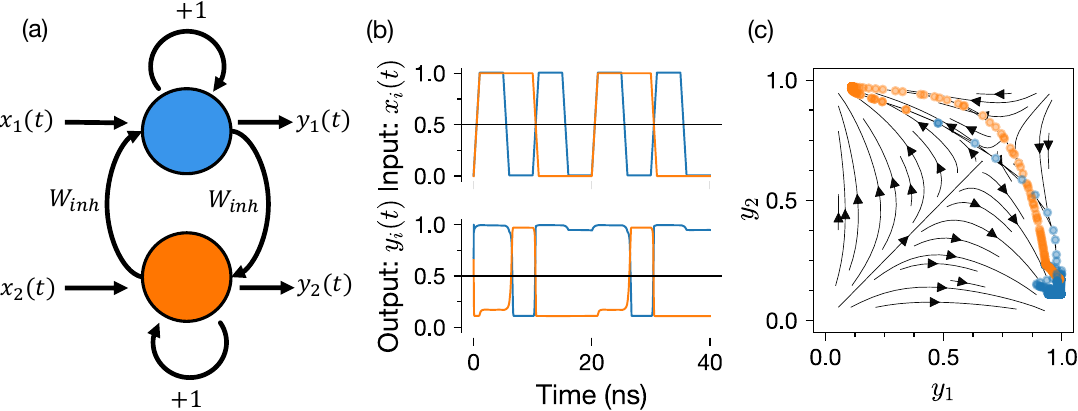}
            \caption{
                (a) Conceptual circuit diagram of winner-take-all simulation showing programmed weights and output signals.
                (b) Transient input and output signals for each of the two neurons.
                (c) State-space plot of outputs, where the colour of the marker corresponds to the neuron with the greater input (the expected winner).
                The vector field depicts the expected dynamics in the absence of any input signal.
            }
            \label{fig:wta-veriloga}
        \end{figure*}

    \subsection{Winner-take-all}
        \label{sec:wta_dynamics} 

        Using the same simulated circuit, by reconfiguring the weights to the matrix Eq.~(\ref{eq:wta_matrix}), we can change the system dynamics.
        The conceptual diagram in Fig.~\ref{fig:wta-veriloga}(a) depict the system with the new weights.
        The mutual inhibitory weight $W_{inh}$ is configured to be -1, so that each neuron strongly inhibits the other.
        For the WTA simulation, we add two more lasers at $\lambda_3$ and $\lambda_4$ for the inputs $x_1$ and $x_2$, which independently perturb neurons 1 and 2.
        Since the neurons are configured identically apart from their weights, the winning neuron is decided completely by which of them has the stronger input.

        Figure~\ref{fig:wta-veriloga}(b) depicts this behaviour.
        The inputs $x_i$ are square waves that implement a truth table.
        Initially, by random chance, neuron 1 (blue) has a slightly higher output, so its activation function saturates (it wins) and it suppresses neuron 2 (orange).
        When $x_1$ goes low and $x_2$ remains high (around \qty{5}{\ns}), the outputs switch.
        When both inputs go low (around \qty{15}{\ns}), the system is already in the basin of attraction around neuron 1 and so it stays there; the system has a memory of its last state.
        Figure~\ref{fig:wta-veriloga}(c) shows these results as a phase portrait.
        The vector field is a plot of Eq.~(\ref{eq:LPF_single_neuron_fb}) using weight matrix Eq.~(\ref{eq:wta_matrix}) in the absence of any input ($x_1=x_2=0$) which shows the two basins of attraction on either side of the line $y_1=y_2$.
        The data points show the trajectory of the system in response to the input perturbations; each point is coloured based on the higher input at that time (the expected winner).
        The high density of orange and blue points in their respective basins of attraction demonstrate the WTA dynamics.

\section{Conclusion}
    \label{sec:conclusion}
    In this work, we have reviewed the dynamics of continuous-time recurrent neural networks implemented in silicon photonics.
    We have explored their fixed points and stability in both feedforward and feedback configurations, and presented a simplified model of how the dynamics of the electro-optic circuit can be mapped to the dynamics of a CTRNN.
    We then demonstrated these dynamics through a few example problems using previously developed Verilog-A based behavioural models, including all relevant parasitic effects present in experimental demonstrations.
    
    The results indicate that there can be significant deviations from the CTRNN model in the presence of these parasitics.
    There is qualitative agreement between the simulated results and the model, but we do not see quantitative agreement.
    We can conclude that while the simulation model and the CTRNN are not isomorphic, they share the important dynamical features of their state space; we can say they have an approximate topological equivalence.
    These results motivate the suitability of silicon photonic neurons for the implementation of recurrent neural networks where the topological behaviour is more important than the quantitative result.
    They also motivate the investigation of programming methods that consider the parasitics to compensate for the differences and improve the quantitative reliability of the system.
    Finally, these deviations support using physical simulation as a tool for predicting and analyzing the results of future on-chip experiments.

    Future work on the Verilog-A based models should focus on three main areas: 1) the implementation of larger scale systems and training of simulated neural networks to implement optimization and control problems, 2) the incorporation circuit schematics for experimental packaging circuits and CMOS control systems, and 3) the integration with chip layout tools.
    Automated generation of SPICE netlists using these Verilog-A models will allow for an interface between the simulator and established neural network packages, such as PyTorch or Nengo.
    Combining these simulations with schematics of packaging circuits will allow for a further level of physical verification between simulation and experiment.
    Finally, integration with layout tools will allow for automated generation of chip layouts based on simulated circuits, bringing the silicon photonics design flow closer to the established electronic design automation process in digital electronics.
    Together, this work will pave the way for experimental realization of large scale silicon photonic recurrent neural networks.

\begin{acknowledgments}
    This work was supported by the Natural Sciences and Engineering Research Council of Canada (NSERC).
\end{acknowledgments}

\appendix

\section{Verilog-A Photonic Neuron Circuit}
    \label{sec:appendix-circuit-diagram}

    \subsection{Photonic Verilog-A Models}
        \label{sec:appendix-veriloga-methods}
        The simulation methodology used to simulate photonic circuits using Verilog-A is covered extensively in \cite{singhNeuromorphicPhotonicCircuit2022}.
        The approach taken is to treat the optical signals as three-element data structures containing the real and imaginary electric field amplitudes along with their wavelength,
        $$
        E = \left( E_{real}, E_{imag}, \lambda \right).
        $$
        This is consistent with the slowly varying envelope approximation.
        This approach discards information about the time-dependant oscillations of the electric field, but it is sufficient for the purposes of this work because it retains information about the optical power and the phase of the signal.
        Treating the optical signals of different wavelengths as distinct data structures allows for the simulation of wavelength-division multiplexed (WDM) systems and is suitable so long as we are not dealing with nonlinear optical effects.

        \paragraph*{Waveguides}
            The waveguide model is assumed to be linear, with a complex index of refraction $\tilde{n}_0=n_0 + i \alpha_0$, to account for phase shift and loss.
            The model converts the real and imaginary parts of the electric field into polar form, applies the loss and phase shift, and then converts back to rectangular form.

        \paragraph*{Phase Shifters}
            The phase shifter model is an augmented waveguide where input voltages and currents can change the values of the real and imaginary parts of the index of refraction.
            Specifically, we use experimentally validated models for the change in refractive index due to either a heater current $I_{heat}$ for the N-doped heater-integrated waveguides, or a modulator voltage $V_{mod}$ for the PN junction microring modulators.
            These are combined into a compound model,
            $$\tilde{n}(I_{heat}, V_{mod})=\tilde{n}_0 + \Delta\tilde{n}(I_{heat}) + \Delta\tilde{n}(V_{mod}),$$
            where only one of the two effects is used at a time depending on the phase shifter being modeled.

        \paragraph*{Couplers}
            The coupler is a four-port model that uses a transfer matrix to calculate the output signals based on the input signals.
            The transfer matrix is assumed to be lossless and is parametrized by the coupling coefficient $k$
            $$
            \begin{bmatrix}
                E_{out,1} \\
                E_{out,2}
            \end{bmatrix}
            =
            \begin{bmatrix}
                t & - i k \\
                - i k & t
            \end{bmatrix}
            \begin{bmatrix}
                E_{in,1} \\
                E_{in,2}
            \end{bmatrix},
            $$
            where $t = \sqrt{1 - k^2}$ is the transmission coefficient, assuming no coupling loss.

        \paragraph*{Microring Resonators}
            \begin{figure}[h]
                \centering
                \includegraphics{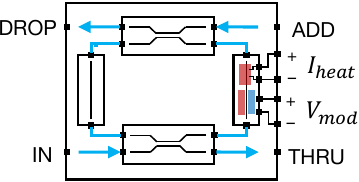}
                \caption{
                    A schematic representation of the Verilog-A model of a microring resonator.
                    The model has four optical ports\textemdash two input ports (IN and ADD) and two output ports (DROP and THRU).
                    The model consists of two directional couplers, a waveguide, and a phase shifter which has four electrical ports for the positive and negative terminals of $I_{heat}$ and $V_{mod}$.
                }
                \label{fig:mrr_schematic}
            \end{figure}         
            
            The MRR model is constructed using the previously described models of the coupler, waveguide, and phase shifter.
            The schematic of the model is shown in Fig.~\ref{fig:mrr_schematic}.
            It is possible to use either the N-doped heater or the PN junction modulator as the phase shifter, depending on which set of electrical ports on the phase shifter are used. 

        \paragraph*{Photodetectors}
            The photodetector model performs the electro-optic conversion from optical power to electrical current.
            The model is constructed from a behavioural current source in parallel with a junction capacitance and a shunt resistor, all in series with a small resistance.
            The electro-optic conversion takes place in the current source, where the current is given by the sum across all incoming optical channels
            $$
                I_{ph}(t) = \eta \sum_i (|E_{real, i}|^2 + |E_{imag, i}|^2),
            $$
            where $\eta$ is the quantum efficiency of the photodetector.

    \subsection{Neuron Circuit Model}

        % circuit diagram figure
        \begin{figure*}[htbp]
        \label{fig:full_circuit}
            \begin{center}
                %\scalebox{0.5}{
                \begin{circuitikz}[american, scale=0.75]
                \draw (0,0)
                to[pD*,l=$\sum_i D_i P_i$] (0,2) % The DROP port source
                to[short] (2,2)
                to[C=$C_p$, -*] (2,0); % The cap
                %to[short, i=$I_{ph}$] (0,0);
                \draw (0,-2)
                to[pD*,l=$\sum_i T_i P_i$] (0,0) % The THRU port source
                to[short, i=$I_{ph}$] (2,0)
                to[C=$C_p$] (2,-2) % The cap
                to[short] (0,-2);
                \draw (2,0)
                to[short] (4,0)
                to[leD*,l=$MOD$, -*] (6,0)
                to[L,l_=$L_{b}$, -] (8,0)
                to[short] (10,0);
                \draw (4,0)
                to[short] (4,2)
                to[short] (6,2)
                to[R=$R_b$] (6,0);
                \draw (6,2)
                to[L=$L_{b}$] (8,2)
                to[isource, l=$I_{bias}$, invert] (10,2)
                to[short] (10,0);
                \draw(6,0)
                to[short] (6,-2)
                to[C=$C_p$] (2,-2); % The cap
                \draw(0,2)
                to[L,l_=$L_{b}$] (0,4)
                to[L=$L_{T}$] (2,4)
                to[vsource, l=$V_{PD}$] (4,4)
                to[short] (10,4)
                to[short] (10,2);
                %++(0,0) node[ground,rotate=90](GND){};
                \draw (0,4)
                to[C=$C_{T}$] (-2,4)
                ++(0,0) node[ground,rotate=270](GND){};
                \draw(0,-4)
                to[L,l_=$L_{b}$] (0,-2);
                \draw(-2,-4)
                to[C=$C_{T}$] (0,-4)
                to[L=$L_{T}$] (2,-4)
                to[vsource, l=$V_{PD}$,invert] (4,-4)
                to[short] (10,-4)
                ++(0,0) node[ground](GND){}
                to[short] (10, 0);
                %++(0,0) node[ground,rotate=90](GND){};
                \draw(-2,-4)
                node[ground,rotate=270](GND){};
                \end{circuitikz}
                \caption{
                    Electrical circuit diagram of a single neuron circuit.
                    The optical interconnects of the system are omitted since these vary for each simulation.
                    Diode symbols are used to represent the Verilog-A behavioural models of photonic components.
                    The balanced photodetector configuration generates the current $I_{ph}$ proportional to the optical power $P_i$ on each channel.
                    The microring modulator $MOD$ is represented using an LED symbol for simplicity.
                    $C_p$ and $L_b$ are the parasitic bond pad capacitance and wire bond inductance, respectively.
                    $R_b$ is the bias resistor.
                    $C_T$ and $L_T$ are the bias tee capacitor and inductor, respectively.
                    $V_{PD}$ is the photodetector bias voltage.
                    $I_{bias}$ is the modulator bias current.
                    $I_{ph}$ is the photocurrent generated by the balanced photodetectors.
                }
            \end{center}
        \end{figure*}
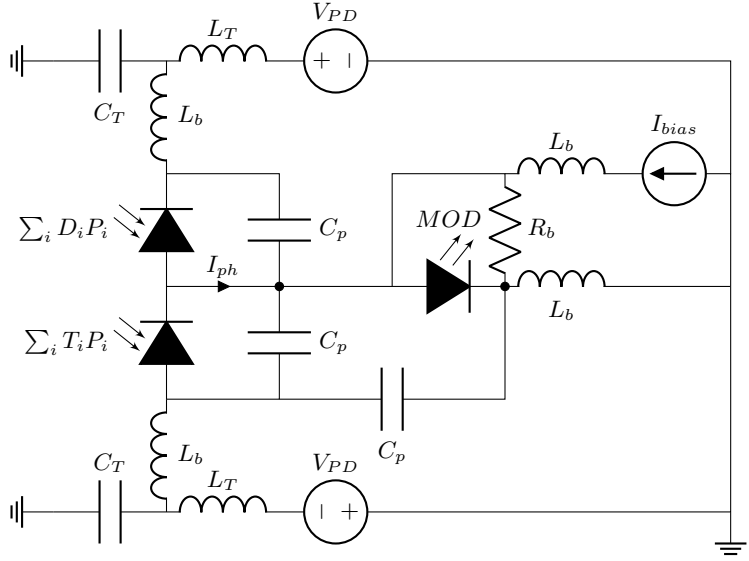
            
        \label{sec:appendix-full-circuit-diagram}
        The electro-optic circuits shown in the main text are simplified circuit diagrams to show the functionality, without the details of the biasing circuits and parasitic components.
        The complete circuit diagram of a single neuron, including the custom Verilog-A models, and active and passive circuit components coming from the typical experimental setup, is shown below.
        The optical waveguides are not shown here.

        The dominant parasitic elements come from the circuit nodes that are connected off-chip using bond pads.
        We model the bond pads as a capacitance $C_p$, and the off-chip wire bond connection as an inductance $L_b$.
        The bond pad capacitance is $C_p=\qty{10}{\femto\farad}$.
        The wire bond inductance is taken to be $L_b=\qty{1}{\nano\henry}$.
        These values are typical for the bond pads and wire bonds.
        
        We also design a simple bias tee made from an inductor $L_T=\qty{1}{\milli\henry}$ and a capacitor $C_T=\qty{2}{\micro\farad}$.
        A bias tee is a three-port circuit that allows separation of the DC and AC components of a signal.
        This is necessary for using off-chip voltage and current sources to bias the photodetectors and the modulator, while still allowing the time-varying photocurrent to remain on chip to modulate the output optical signal.
        This bias tee design is far from optimal and does not use realistic component selection, but is functional for the signals of interest in this work.

\section{Weight Mapping Procedure}
    \label{sec:appendix-weight-mapping}
    
    In order to implement an arbitrary weight vector using a bank of microring resonators, it is necessary to perform a characterization of the microrings for a given channel (carrier wavelength) and ring radius of interest.
    In this Appendix, we demonstrate feedforward characterization of a 5-channel MRR weight bank.
    We choose here the base radius of the first ring in the weight bank to be $R_0=\qty{8}{\micro\metre}$ and increment the radius for each ring by $\delta R=\qty{12}{\nano\metre}$.
    Similarly, the carrier wavelengths of the first channel are chosen to be $\lambda_0=\qty{1548.7}{\nano\metre}$ and increment the wavelength by $\delta \lambda=\qty{2.35}{\nano\metre}$.

\begin{figure}[h]
        \centering
        \includegraphics[width=\linewidth]{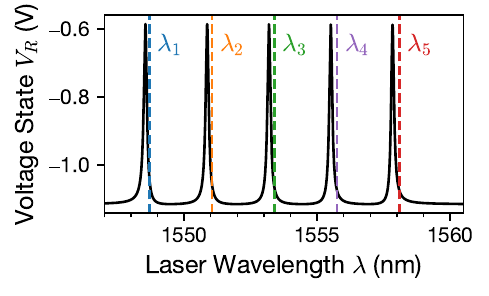}
        \caption{
            Spectrum of voltage measured at the neuron junction for a weight bank with $R_i=R_0+i\delta R$.
            The vertical lines correspond to the chosen channels at $\lambda_i=\lambda_0 + i\delta \lambda$.
        }
        \label{fig:junction_voltage_spectrum}
    \end{figure}

    After constructing a neuron as in Fig.~\ref{fig:photonic_neuron} with these parameters, we set the heater current for each ring to 0, sweep the input wavelength and measure the voltage across the modulator and obtain the spectrum in Fig.~\ref{fig:junction_voltage_spectrum}.

    \begin{figure}[h]
        \centering
        \includegraphics[width=\linewidth]{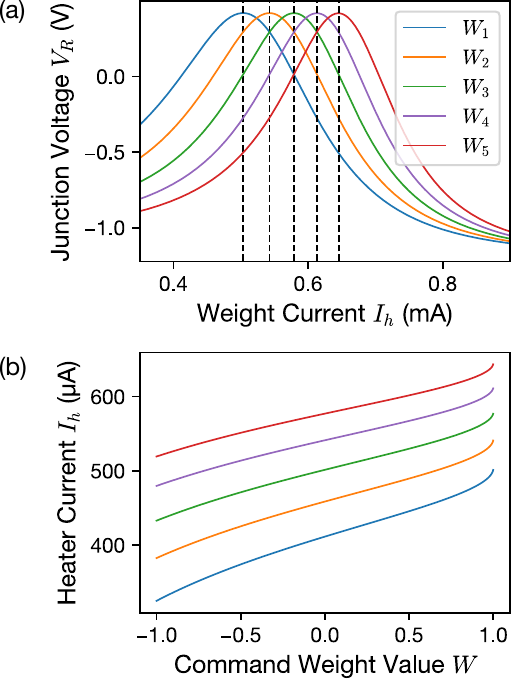}
        \caption{
            (a) Junction voltage response to a sweep of weight tuning current for each channel.
            (b) Final weight mapping curve for each weight channel.
            For any desired weight $-1<=W_i<=+1$, we have a unique heater current that we can apply to achieve this.
        }
        \label{fig:weight-mapping}
    \end{figure}

    Next, for each channel independently, we can set the pump laser to $\lambda_i$, and sweep the input heater current from \qtyrange{0}{1}{\milli\ampere} and again measure the voltage at the neuron junction to obtain the individual resonance curve of each ring.
    We can then determine the value of the applied current corresponding to bringing the ring onto resonance with the given channel, as seen in Fig.~\ref{fig:weight-mapping}(a).
    This is the current for each channel that corresponds to the most light being coupled to the DROP port.
    Note the asymmetry in the response; there is no bias current supplied and yet most of the resonance curve lies below \qty{0}{\volt}.

    Finally, we discard all the data above the resonance current, truncate the curve so that it is symmetric about \qty{0}{\volt}, and invert the relationship between heater current and junction response.
    The final result can be seen in Fig.~\ref{fig:weight-mapping}(b), where we have now normalized the response so that the maximum voltage corresponds to $W=-1$ and the minimum corresponds to $W=+1$.

    It is important to note that this weight mapping procedure is the theoretically optimal way to map heater currents to MRR weights, and is fit for the purpose of simulation, but it does not account for many of the important physical effects that limit the precision in experiment.
    Fabrication variation, environmental temperature fluctuations, and photoconductivity of N-doped heaters are some of the effects that are not yet accounted for in the model and these are left for future work.

% force bibliography to be on a new page
% \clearpage 

%\nocite{*}

\typeout{}
\bibliography{refs} % Produces the bibliography via BibTeX.

\end{document}